Preprint version January 18, 2017Title: **Non-linear temperature-dependent curvature of a Phase Change Composite Bimorph beam**

Author:
*Prof. Greg Blonder*
*Mechanical Engineering Department*
*Boston University*
*Room 202F*
*730 Commonwealth Ave*
*Boston MA 02215*
*gblonder@bu.edu***Keywords**: bilayer; phase-change; microspheres; thermal curvature

**Graphical Abstract**

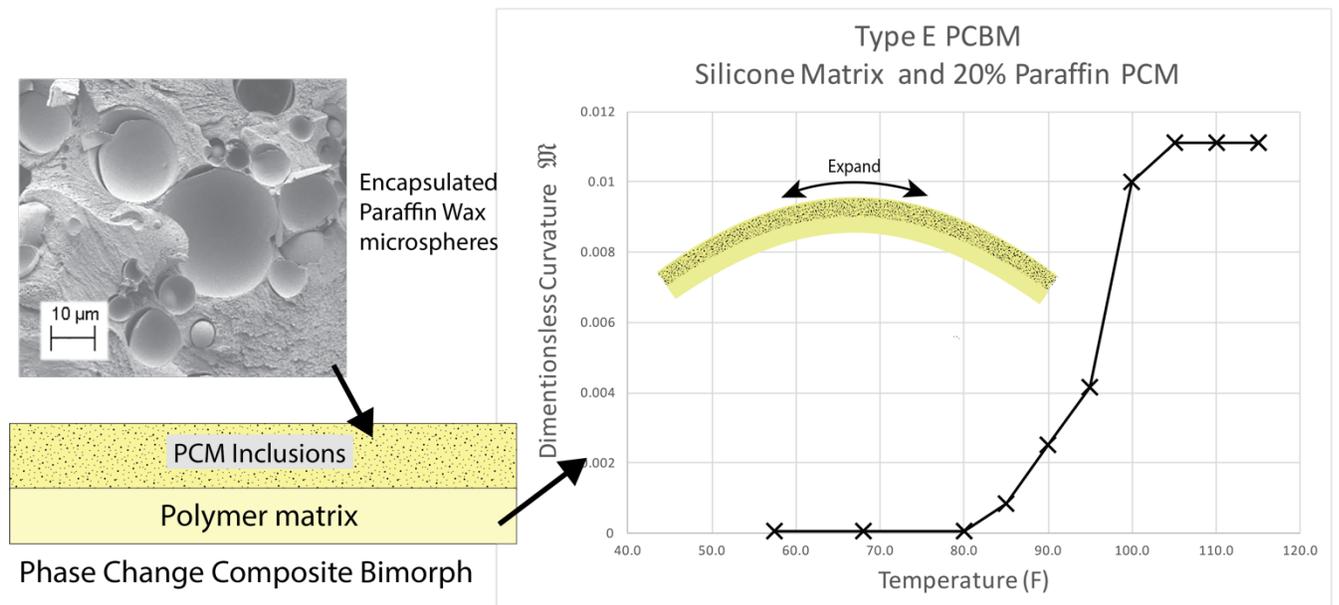

**Abstract**: Bimorph films curl in response to temperature. The degree of curvature typically varies linearly with temperature and in proportion to the difference in thermal expansion of the individual layers. In many applications, such as controlling a thermostat, this gentle linear behavior is acceptable. In other cases, such as opening or closing a valve or latching a deployable column into place, an abrupt motion at a fixed temperature is preferred. To achieve this non-linear motion, we describe the fabrication and performance of a new bilayer structure we call "Phase Change Composite Bimorph (PCBM)". In a PCBM, one layer in the bimorph is a composite containing small inclusions of phase change materials. When the inclusions melt, their large (generally positive and >1%) expansion coefficient induces a strong, reversible step function jump in bimorph curvature. The measured jump amplitude and thermal response is consistent with theory.Page 1 of 17   Blonder PCBM

## 1. Background

Bimorph mechanical structures are ubiquitous. Even though a bimorph's intrinsic temperature sensitivity is low, with a large enough temperature difference and a high enough aspect ratio, large bending motions are feasible.

For example, coiled bimetallic films are the de-facto technology of choice in hundreds of millions of thermostats. Thermostatic faucets are required by code to prevent scalding. Heater-wrapped bimetallic strips are still blinking away in automobile turn flashers.

However, there are many situations where a standard bimorph film is inadequate. In biomedical devices, only a few degrees separate core from skin temperatures. This temperature difference may be too small to trigger a bimorph control. Additionally, since the bimorph's response is linear in temperature, there are situations where temperature prematurely warps the bimorph, causing a valve to leak rather than seal.

Here, we offer an alternative to the standard linear bimorph. Our structure, which incorporates a phase-change material into a bilayer, remains thermally matched at all but the transition temperature. When it does curl, only a few degrees of temperature difference produce a jump in curvature that normally would require 5-10 times the thermal excursion in a standard bimorph.

A non-linear bimorph that "switches" state at a fixed temperature should enable many new sensors, actuators and applications.

## 2. Linear Bimorph Theory

In any structure with two or more dissimilar layers, temperature changes will create differential strain or motion. When that motion is intentional, we call the bilayer a *bimorph*.

For a conventional bimorph beam in the linear approximation, curvature is proportional to the difference in thermal expansion coefficient between the two layers.

At one unique temperature determined by processing conditions, the film lays flat; above and below it curls in opposite directions.

The rudimentary bimorph bending equation (determined entirely geometrically by matching each layer's thermally expanded length and ignoring bulk strains), is well established. For thin films, the deflection of a cantilevered beam tip "d" (or the radius of curvature "r") is approximately:

$$d \cong \frac{\Delta\alpha \, \Delta T}{\delta} L^2 \qquad (1)$$

$$r \cong \frac{\delta}{2 \, \Delta\alpha \, \Delta T} \qquad (2)$$

"L" is the beam length, "$\Delta\alpha$" is the difference in thermal expansion coefficient of the two layers, "$\Delta T$" is the temperature difference relative to the layflat temperature, and "$\delta$" is the thickness of the beam.

For example, consider a high expansion polyethylene (PE) film laminated to a thin low expansion sheet of polyethylene terephthalate (PET), with a $\Delta\alpha$ of 150 ppm/C, a thickness $\delta$ of 4 mils (0.01cm) and a length > 10 cm. When heated 20C above its layflat, this strip will curl into ~3 cm diameter cylinder. Due to the low mass of the film, response times are of order a second. This response may be compared to the more common bi-metallic strip. In this case, $\Delta\alpha$ is closer to 5 ppm/C and the films tend to be thicker to avoid kinking. So the curvature is 100x smaller than with an all-plastic bimorph.

In a more complete model of a bimorph, the layer's relative stiffness matters. Rearranging the terms of Timoshenko's bimetal Thermostats equation, [1] the radius of curvature can be written as:

$$r = \frac{t_1 + t_2}{2\Delta\alpha\Delta T}[1 + \beta] \qquad (3)$$

and we define the dimensionless curvature[1] as

$$\mathfrak{M} \equiv \frac{t_1 + t_2}{r} = \frac{\delta}{r} = \frac{2\Delta\alpha\Delta T}{1 + \beta} \qquad (4)$$

$\beta$ is the Young's modulus correction to a simple bimorph cantilever:

$$\beta = \frac{1}{3} \frac{E_1 t_1^3 + E_2 t_2^3}{(t_1 + t_2)^2} \left( \frac{1}{E_1 t_1} + \frac{1}{E_2 t_2} \right) \qquad (5)$$

The radius is minimized (e.g. most curled) when

$$\frac{E_1 t_1^2}{E_2 t_2^2} = 1 \qquad (6)$$

---

[1] In honor of Timoshenko, $\mathfrak{M}$ is similar to the cursive form of the Russian letter "T".

When $E_1=E_2$, both film's thicknesses should be identical to maximize curling. If one film is much thinner than the other, that thin film should be very stiff. For example, if $t_1$ is 3 mils of stainless steel, and $t_2$ is a 10x thicker polymer, then Film 2's $E_2$ should be 200GPa/100= 2GPa. This is consistent with nylon or polycarbonate, but not low density PE.

In Fig. 1, plotting $\beta$ vs the Young's Modulus ratio $E_1/E_2$, for various film layer thickness ratios $t_1/t_2$, we find:

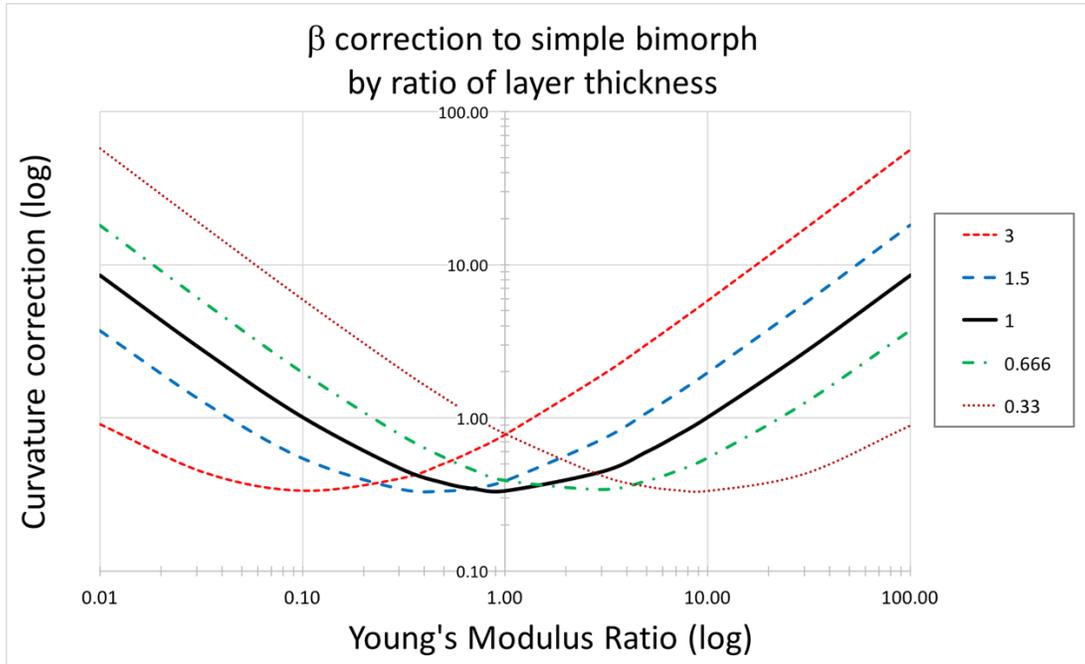

Fig. 1 Plot of the bilayer beam curvature correction $\beta$ [ Eg. ( 5)] as a function of the layer's Young's modulus ratio, for bilayers with different layer thickness ratios. As Timoshenko noted, there is a factor of 10 range in relative modulus over which the curvature is relatively constant.

Note the lower bound for $\beta$ is 1/3$^{rd}$. The minimum is also very shallow, and barely shifts even with a factor of 10 in relative modulus. Generally, stiffer materials are also lower expansion, so given the broad minimum, if strong curling is the goal, reducing the stiffness in exchange for a higher $\Delta\alpha$ is a sensible tradeoff.

## 3. Non-linear bimorphs

In many cases, the bimorph's symmetric temperate response (negative or positive curvature below or above the layflat temperature) can be problematic. For example, in the case of an "instinctive" air vent to control a hot-air damper [2], a food-packaging [3] inspired bimorph film based on the ubiquitous mylar/PE potato chip bag, acts as a flapper valve. Below the trigger temperature it should lay flat sealing the vent. Yet open wide at temperatures above. Unfortunately, the continuous curvature with temperature makes tight sealing problematic.

One way to introduce a non-linear effect is with a phase transition to supplement the material's bulk expansion coefficient. For example, a wax thermostat in an automobile's cooling system's is powered entirely by a liquid/solid volume change, opening sharply at a preset temperature. Volume changes on melting can be very large- up to +15% in paraffin, +10% for stearic acid, -3% for gallium, and -8% for ice.

Assuming a robust process can be developed to incorporate a Phase Change Material (PCM) into one of the films, the bilayer would experience a rapid jump in curvature through the phase transition. We call this a PCBM for "Phase Change Composite Bimorph".

Defining $\acute{\alpha}$ as the fractional jump in length of the PCM layer through the transition (which plays the role of $\Delta\alpha\Delta T$ in (4)) we see that

$$\mathfrak{M} = \frac{2\acute{\alpha}}{1 + \beta} \qquad (7)$$

Oriented thin film vanadium oxide [4] has been studied as a PCBM in MEMs devices. But vanadium oxide's bulk expansion is basically zero, and the material is expensive. In this work, we seek to exploit an extreme bulk volume change in a homogeneous thin film.

To harness the difference in volume, the liquid phase must be contained, and in tight contact, with the surrounding matrix. Various combinations of PCMs and matrixes were investigated, including waxes such as stearic acid or bayberry, and polymers such as high durometer silicones and urethanes, or absorbing sponges or films soaked with PCMs. Not all PCMS are compatible with all matrixes. For example, wax prevents epoxy or urethane from completely curing, and sometimes even generates gas bubbles by chemical reaction. Other polymers may delaminate/crack at the interface between the layers, due to high shear forces produced at the phase transition.

A particularly convenient and versatile approach is enabled by PCM paraffin microspheres (µPCM). Here, PCMs are encapsulated by a thin, melamine-formaldehyde shell. Microsphere shells block any deleterious chemical interactions between the PCM payload and the surrounding matrix. The shell also prevents the enclosed paraffin/hydrocarbon oil mixture from evaporating.

µPCM powder is commercially available as a latent heat storage media [5] [6] [7] or for smart clothing [8] (*Microtek MPCM37D dry powder, Dayton OH is used throughout this paper. The paraffin wax/hydrocarbon PCM is proprietary. Various transition temperatures are available. Other manufactures employ acrylic shells.*)

Standard PCM microspheres are carefully designed to avoid expansion coefficient stresses on the shell or matrix, generally by curing the thin plastic shell around the melted, high volume PCM phase (see Fig. 1(c) for a cross-section of µPCM in epoxy). The PCM may comprise over 85% of the total volume of the microsphere. Depending on shell thickness, it will either deflate below the melting point, or trap an air bubble inside.

Air buffers any expansion of the PCM on melting and would limit transmission of its volume change to the surrounding matrix. However, when the shells are thin, and the matrix is cured <u>below</u> the phase transition temperature, the shells collapse (as confirmed by SEM). Curing <u>above</u> the phase transition would lock-in the expanded microsphere shell and preserve the entrained air bubble.

The base layer, which does not contain PCMs, should ideally match the matrix in thermal expansion coefficient (in fact, it may be the identical composition as the matrix). Expansion coefficient matching assures the film remains flat up to the PCM phase transition temperature.

As measured by their Archimedes displacement in ethanol, the 100F phase transition microspheres used in this study expand by around 9% in volume through the melting point.

In our experiments, the μPCM is blended into a matrix as small spherical inclusions into one layer of a bimorph *Fig. 3*. According to Turner [9], the effective expansion coefficient for an isotropic mixture of two different materials can be approximated as:

$$\alpha_{effective} = \frac{c_1 \alpha_1 \kappa_1 + c_2 \alpha_2 \kappa_2}{c_1 \kappa_1 + c_2 \kappa_2} \qquad (8)$$

where

$$\kappa_1 = bulk\ modulus\ of\ material\ 1$$
$$c_1 = concentration\ material\ 1$$
$$\alpha_1 = expansion\ coefficient\ material\ 1$$

In other words, the jump ά would be leveraged by the PCM concentration and relative modulus. For example, if the PCM generates a 10% volume change at the transition, with a concentration around 10%, the volume jump would be 1%, implying a linear expansion ά ~ 1/3$^{rd}$ %. Thus, we expect to measure changes in $\mathfrak{M}'s$ of around 0.003.

To confirm their suitability for this application, μPCMs were mixed into epoxy at the 20 wt% level, and cast into 10cm long rods. As seen in Fig. 2, the rod's length abruptly rises at the 100F phase transition. Note the jump in length at 100F is around 0.5%, roughly consistent with a 20% MCM concentration and a linear expansion of 1/3$^{rd}$ of the 9% PCM volume expansion.

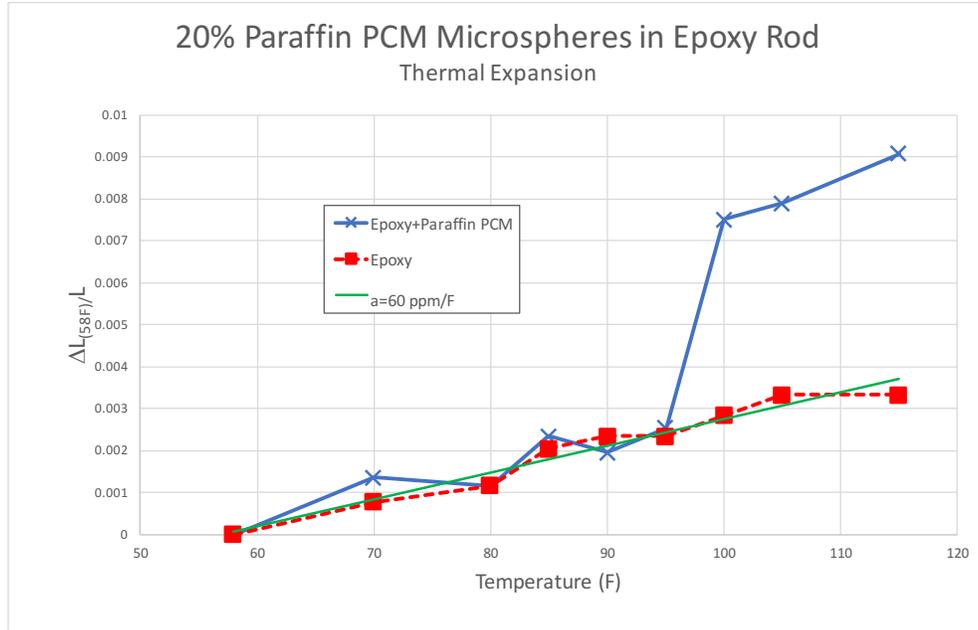

*Fig. 2* Expansion of an epoxy rod containing 20% by wt. µPCM inclusions. The epoxy has a bulk linear thermal expansion coefficient of 60 ppm/C.

## 4. Laminated PCM Structures

In this paper, three different PCBMs geometries were investigated.

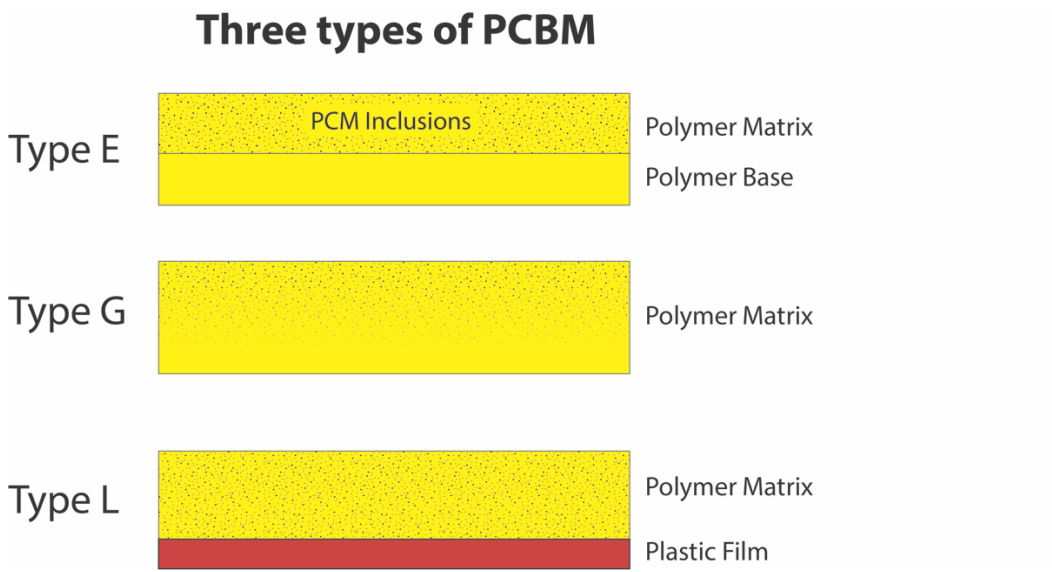

Fig. 3 Three different PCBM structures fabricated in this study. The small black dots represent the µPCM inclusions in the composite.

**Type E (Epi):** Two layers cured from the same polymer- one layer containing a uniform distribution of µPCM. Since both layers are fabricated from the identical polymer, differential expansion is a purely a response to the temperature dependence of the embedded µPCM.

**Type G (Graded):** Before the polymer matrix cures and while its viscosity remains low, the µPCM inclusions either sink (or float) to one surface, depending on their specific gravity relative to the matrix. Since the matrix's expansion coefficient is common to the high and low PCM concentration regions, any temperature induced bending is a function of the µPCM phase transition alone.

**Type L (Laminated):** A layer of polymer uniformly filled with µPCM is either cast on a thin plastic film, or later bonded to a separate plastic film. As the expansion coefficients generally differ between the plastic and polymer, the laminated film will curl in response to both the difference in film expansion coefficients, as well as the PCM transition. It may also undergo a buckling instability [10].

PCBMs were constructed from a phase change material, a compatible polymer matrix, and an assembly protocol to create a bilayer. All of the PCBM samples were compression molded between ½" thick flat aluminum plates. Each plate was covered with a 5 mil release sheet- either PE or PET, as none of the polymers used in this study sticks to PE or PET, post-cure. A parallel gap between the two release sheets is maintained by removable spacers.

To mold the samples, one release sheet is placed on the bottom aluminum plate. Spacers, around 1 mm high, are arranged around the perimeter. A glob of curing polymer, typically an epoxy with the consistency of peanut butter (30k-100kCP) is placed in the center. Then the second release sheet and upper aluminum plate is arranged on top. Normally, the ½" thick plate provided sufficient pressure to flatten the polymer glob- if not, additional weight was applied.

Aluminum plates act as a thermal heat sink, reducing the chance that an exothermic cure reaction might drive the PCM through its phase transition.

After the polymer fully cures, the compression mold is disassembled and the release sheets peeled away from the hardened polymer disk.

Most epoxies have a relatively low (35-45C) glass temperature. During the transition from glassy to elastomeric behavior, coefficients of thermal expansion may increase by 3-5x. This rise might confound measurements of the phase transition jump $\acute{\alpha}$. Thus, high glass temperature (>80C) epoxies were chosen to encapsulate low temperature 100F (37C) microspheres.

In addition, µPCMs typically agglomerate into much larger granules post formulation, often 10-100x greater in diameter than the spheres themselves. If these agglomerates are cast into a high surface tension matrix, the interstitial air will buffer any volume expansion. However, agglomerates can be dissipated by a combination of vacuum de-airing the filled-uncured matrix, choosing a matrix resin that has a low surface tension with the shell, and mechanical shearing.

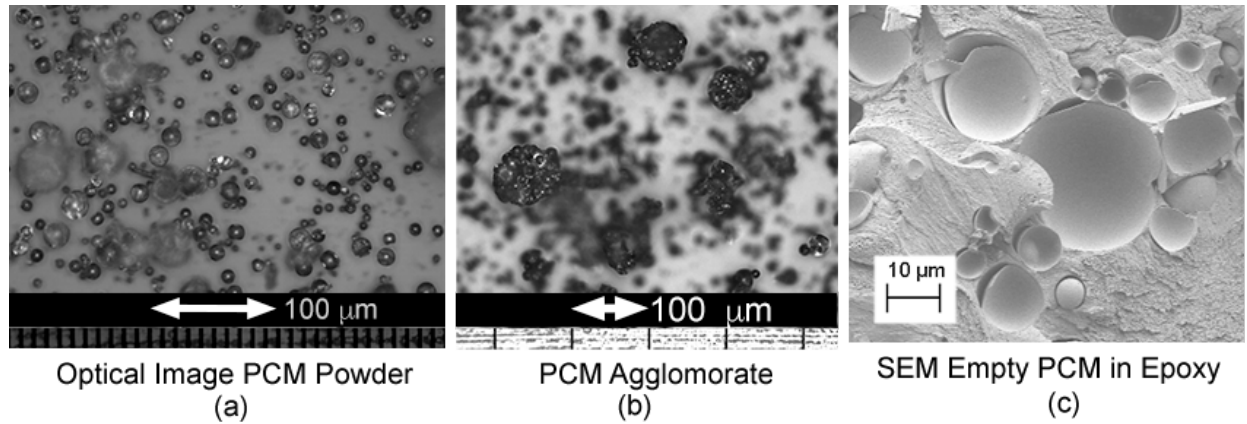

*Fig. 4 Micrographs of paraffin wax microspheres. The µPCM powder consists of individual spheres (a), average diameter 10 µm, and agglomerates of spheres (b). The epoxy matrix fully infiltrates the agglomerates, as evidenced in the cleaved SEM cross-section (c).*

## 5. Type E µPCM/silicone matrix fabrication and results

A convenient, air-drying silicone putty is Sugru (*FORMFORMFORM, Hackney, East London*). 20% by wt µPCM was kneaded into 5 gms of Sugru, then compression molded into a flat 0.75 mm thick disk between two sheets of PE. A similar disk was compression molded out of virgin Sugru. One sheet of PE was peeled from each disk, and the two free, tacky surfaces laminated together under pressure between plates with a 1.5 mm spacer. Finally, both layers of PE were peeled away and the PCBM allowed to air dry for 24 hrs on a fine mesh screen to encourage air-flow.

This 6 cm diameter PCBM disk was mounted on a stand engraved with fiducials indicating the radius of curvature. The stand was placed in a temperature controlled air or water bath which scanned the temperature from 60F to 125F over a period of 30 minutes. Film curvature was measured optically by analyzing still images with custom MATLAB code.

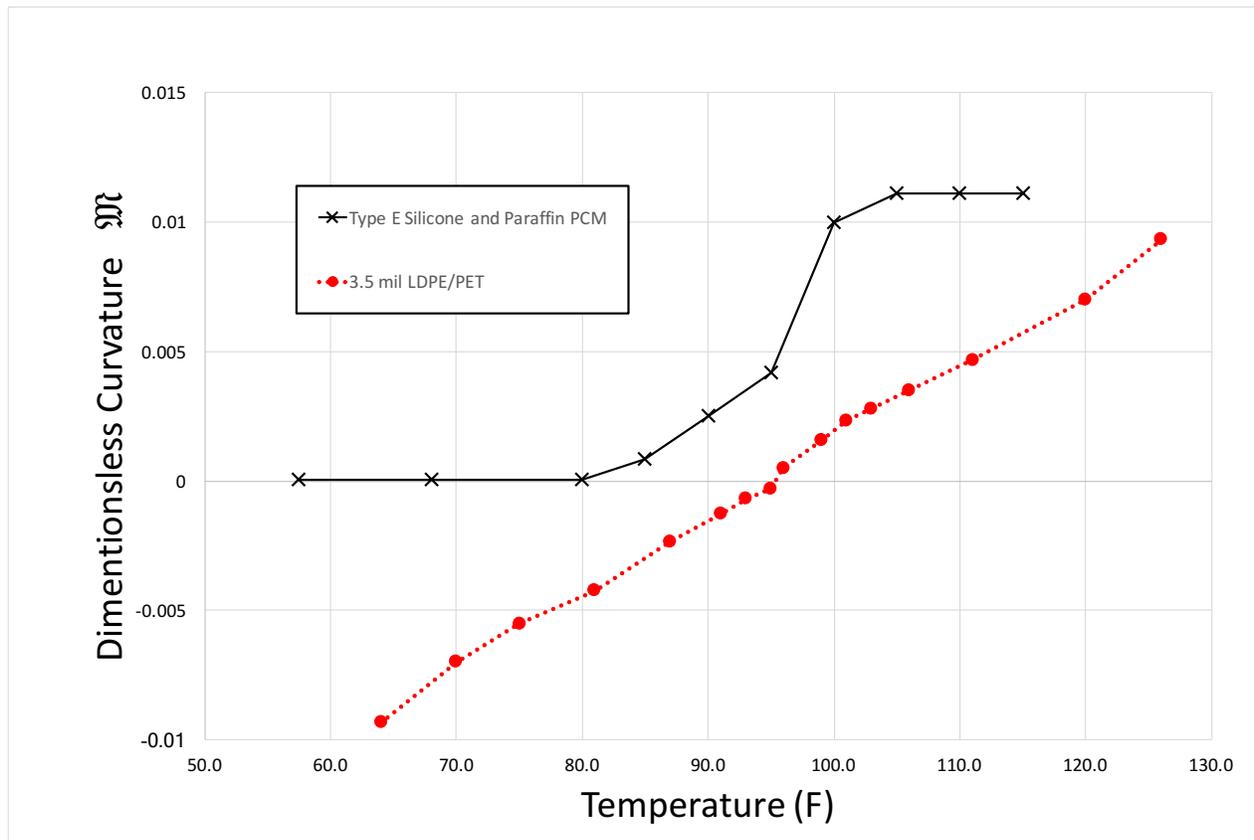

*Fig. 5. Dimensionless curvature vs temperature for a 1.5 mm thick Type E silicone matrix/paraffin wax microsphere PCM, compared to an all-plastic bimorph (3 mil thick film of low density polyethylene laminated to 0.5 mil PET).*

For comparison purposes, this non-linear jump is compared in Fig. 5 to the behavior of a "standard" thermal bilayer- a 3.5 mil thick low density polyethylene/PET structure normally intended for packaging coffee (dotted line in Fig. 5). As expected, this "standard" film demonstrates a completely linear variation in curvature with temperature.

The abrupt jump near the phase transition seen in Fig. 5 is reversible without hysteresis. However, the transition begins ~10F below the nominal melting point because the paraffin wax mixture itself exhibits a range of melting points. Never the less, the general principle and jump amplitude is consistent with the simple theory above.

**6. Types E, L and G epoxy/µPCM matrix results**

Dozens of Type E, L and G films, all between 0.5mm and 2mm in thickness, all with epoxy as the matrix, all containing 20% by wt. µPCM in one layer, were fabricated according the methods described in the Appendix.

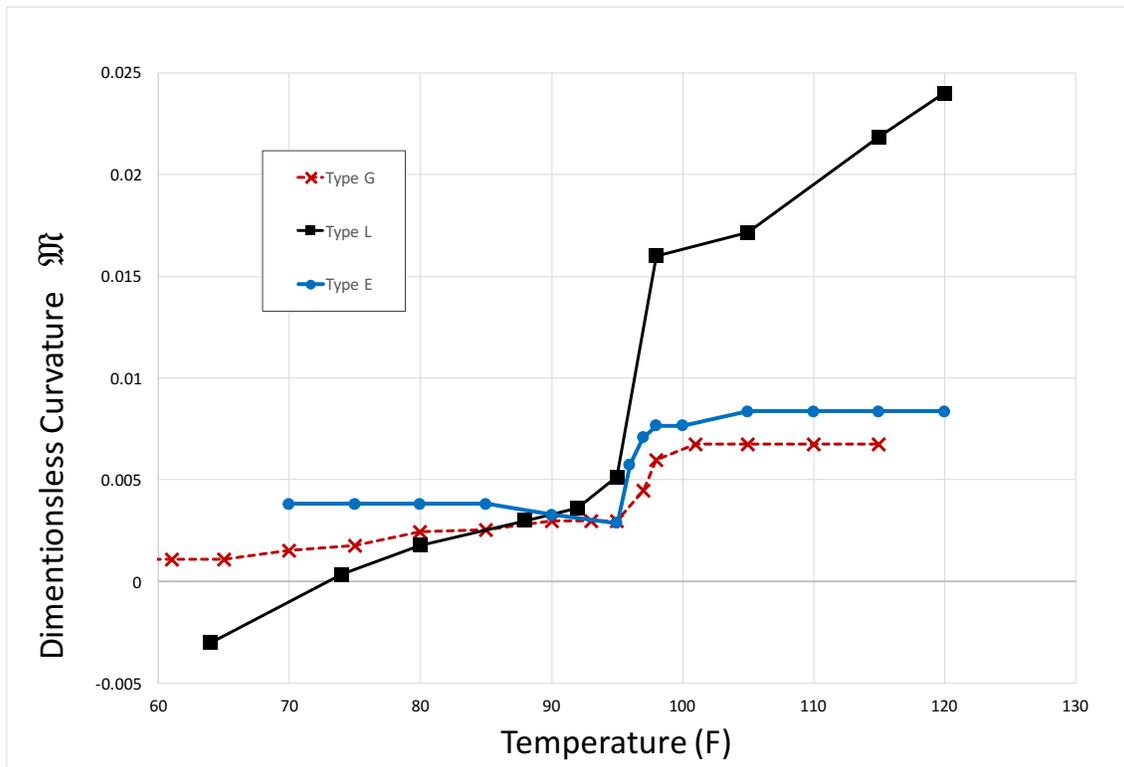

*Fig. 6 Types E, G and L structures compared.*

The Type G and E films, despite the radical difference in structure, behaved similarly, with a more-or-less constant curvature except at the 100F phase transition.

Note how the Type L phase change jump of Fig. 6 is superimposed on the linear background of a standard plastic bimorph. Depending on thickness and stiffness, the film would also buckle when driven from positive to negative curvature. [8] [11]

We also tested unencapsulated gallium inclusions (See appendix). Gallium expands on cooling. Thus the dimensionless curvature in Fig. 7 *declines* as the temperature *rises* through the melting point.

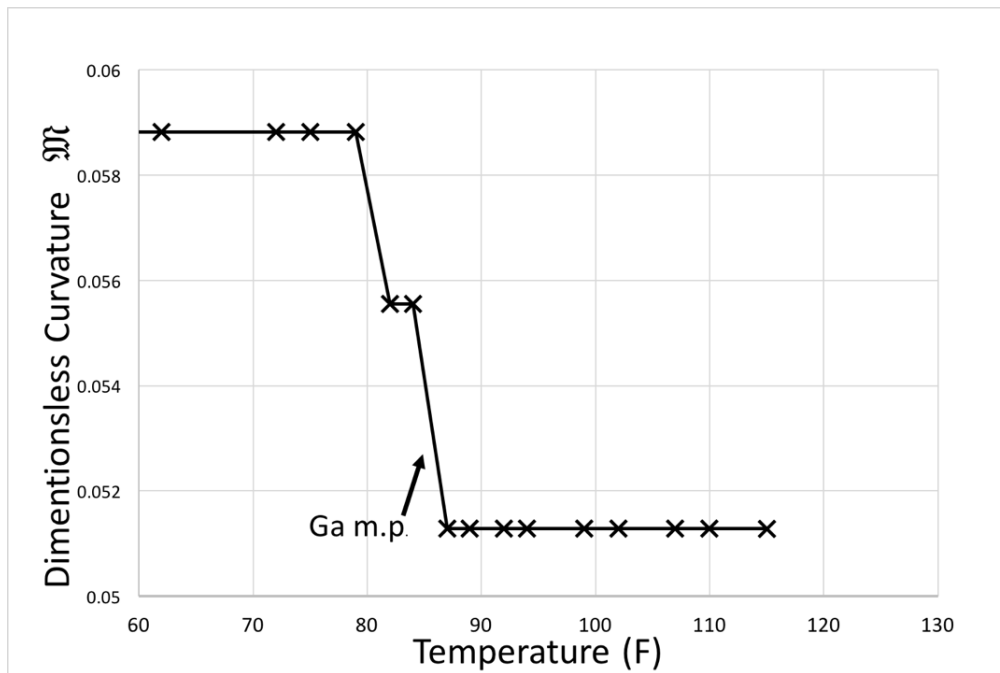

Fig. 7 Type E Epoxy/ 5% by wt. Ga PCBM. Gallium inclusions range in size from 0.5 to 50 µm. Film thickness 60 mils.

## 7. Discussion

These rudimentary bimorph structures clearly demonstrate the basic PCBM principle. By supplementing a film's uniform thermal expansion with a discontinuous jump in volume at a phase transition, a new world of applications, unavailable to standard bimorphs, are now accessible.

A number of design improvements to these initial structures should be investigated. Clearly, the reliance on microencapsulation or small inclusions limits both the magnitude of the effective volume change, and the range of compatible materials.

A resin that segregates on curing into a matrix and a non-percolating PCM phase, may simplify the assembly process. Homogeneous phase change materials which undergo a solid-solid transition, and match the underlying matrix mechanical properties, are potentially superior. But we have yet to identify a practical candidate. Additionally, combining the PCBM with mesoscopic fabrication techniques, such as embossing, lithography or laser cutting, may help control the direction and extent of motion. This would enhance its capabilities as an actuator or sensor.

A PCBM combines many bimorph attributes with those of a two-way shape memory alloy. While the maximum strain and energy/volume is lower, the PCBMs are inexpensive, versatile, non-metallic and require no thermal training. They are well suited for high volume, large area applications.

We speculate that PCBM structures also exist in nature, and may play a role in geological processes such as frost heaving or spalling, or perhaps warp biological structures in response to external temperate triggers. For example, the difference in motion between a waxy leaf cuticle and the underlying moist dermis, may help leaves close in winter. Or, supplementing the accepted "ice lens" [12] theory of soil frost heaving, a layer of saturated forest litter may create a PCBM and bow into an open arch as the cold air crystalizes the outer surface into ice.

## 8. Conclusion

A new bilayer composite thin-film beam structure is described. This structure incorporates a bulk phase change material as small inclusions in one layer of a bimorph. The resulting structure curls abruptly, and reversibly, at the phase transition temperature. Large curling and effective expansion coefficients are demonstrated. Such PCBMs might be employed in various self-assembly mechanisms and actuators, and represent a new modality beyond the range of simple bimetallic strips and shape-memory alloys.

**Acknowledgments**: This exploratory research was funded by the author.

**Appendix**

**Type E µPCM/epoxy film recipe:** A thick, room temperature curing, high-viscosity epoxy was chosen for its relatively high glass temperature and ease of molding (*MG Chemicals Burlington Ontario Canada 832HT. Viscosity 46,000 Cp, glass temperature 68C*).

Microtek MPCM37D powder at the 20 wt % level was mechanically mixed into the epoxy and resin. A 10 gm portion was compression molded between 30 mil spacers and cured for 24 hrs.

After curing, one layer of polyethylene was peeled off. The epoxy surface was lightly sanded, then an additional 10 gm of epoxy dispensed, this time without PCM. 60 mil spacers and a new layer of polyethylene sheeting completed the sandwich- it was then compression molded and cured as above.

Testing was delayed for an additional 72hrs to allow the epoxy to completely cure at room temperature. Gently curing below the 100F phase transition is critical to lock in the low volume phase, and to collapse air bubbles within the microspheres.

**Type G µPCM/epoxy film recipe:** A low viscosity catalyst and resin epoxy (*Atom Adhesives F113 Fort Lauderdale, FL , viscosity ~200Cp, glass temperature ~100C*) was thoroughly mixed together, then 20% by wt. µPCM microspheres were blended in and de-aired for a half hour under vacuum. At this point the viscosity had increased to around 500 Cp. This epoxy was selected for its 3-4 hour pot-life at room temperature, and relatively high glass temperature.

The composite mixture was compression molded between 5 mil PET sheets with 50 mil spacers.

In a low Reynold's number environment, the classic Stokes terminal velocity is

$$u_\infty = \frac{1}{16}\frac{gd^2(\rho_e - \rho_s)}{\mu_e} \qquad (9)$$

where g is the gravitation acceleration, d the microsphere diameter, $\rho_e$ the epoxy density, $\rho_s$ the microsphere density, and $\mu_e$ the epoxy viscosity (~ 1Pa-s). The epoxy density is 1200 kg/m$^3$, while the microspheres are 900 kg/m$^3$, so they float to the surface. Microsphere diameters (as measured by SEM) average around 10 µm, but range from 3 µm individual spheres, to 100 µm clusters. Buoyancy and differential drag will size-separate the spheres as they rise. $u_\infty$ for the clusters is around 10$^{-3}$ mm/sec; after one hour they will traverse the 1mm film thickness, forming a dense floating mat.

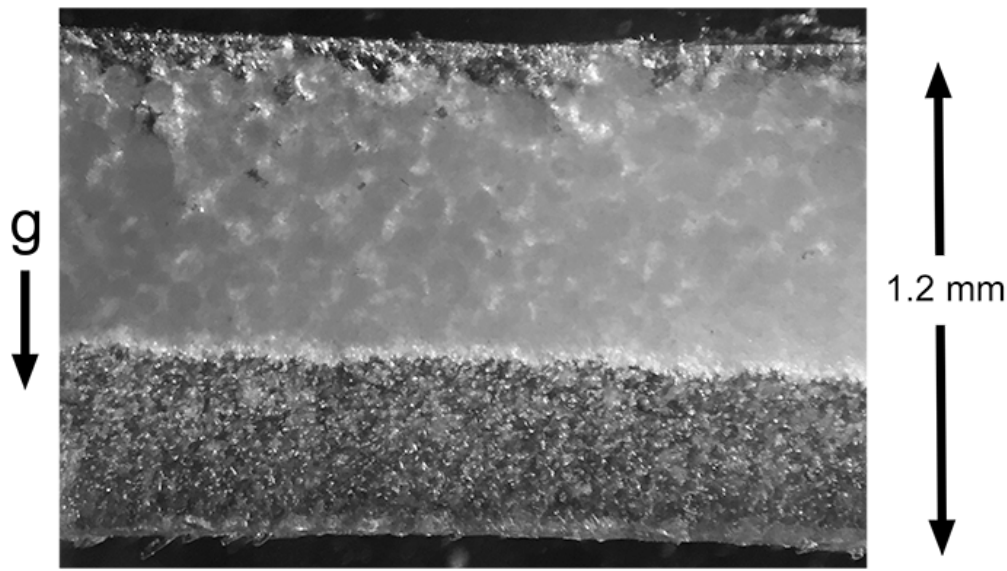

*Fig. 8 Type G PCBM, epoxy matrix. Dark-field photomicrograph. The bright spots at the bottom of the film are individual 10 μm μPCM particles, the larger white clumps are clusters of μPCM powder.*

**Type L PET/epoxy+μPCM film recipe.** A thick, room temperature curing, high viscosity epoxy was chosen for its relatively high glass temperature and ease of molding (*MG Chemicals Burlington Ontario Canada 832HT. Viscosity 46,000 Cp, glass temperature 68C*).

Microtek MPCM37D powder at the 20 wt % level was mechanically mixed into the epoxy and resin. A 10 gm portion was compression molded between a 3 mil sheet of PET and a 5 mil sheet of PE. To improve adhesion, the PET film was sanded and treated with hexamethyldisilazane.

After curing for 24 hrs the PE sheet was peeled off the PCBM disk. Excess PET was removed with a scissor. Testing was delayed for an additional 72hrs to allow the epoxy to completely cure at room temperature.

**Type E gallium - epoxy recipe:** Gallium shrinks on melting, so this Type E film must be mixed and cured above the melting point to lock-in the low volume phase. For that reason, all fabrication steps took place in a 100F oven.

Warm epoxy resin and catalyst (*Devcon H2, Danvers, MA*) was mixed together, and then 5% by vol. liquid gallium was blend in using a high speed motor attached to a smooth 3/8" roller. The shearing forces between the roller and the container, along with the high epoxy viscosity created a fine, gray emulsion of gallium in epoxy. Spherical Ga droplets covered a wide range of diameters averaging 10µm. This composite was compression molded between two 5 mil PET sheet held apart 1mm spacers.

After curing, one layer of PET was peeled off. The surface was lightly sanded, then an additional 10 gm of epoxy dispensed, this time without gallium. 60 mil spacers and a new layer of polyethylene sheeting completed the 8 cm diameter sandwich- it was then compression molded and cured as above.

Testing was delayed for an additional 72hrs to allow the epoxy to completely cure at room temperature.

The phase transition jump begins 10F below the nominal 86F melting point of gallium. In the case of gallium, the melting point is depressed [13] in small submicron particles. Other phases [14] may also be present which melt at a lower temperature.

Gallium is also noted for its enormous undercooling [15], especially when constrained by a smooth surface like a wall of cured epoxy. To assure gallium inclusions were solidified before temperature scanning, the Type E disks were placed on dry ice for one hour. Then a temperature scan from 60F to 120F in a water or air bath was performed. Scanning from 120F to 60F exhibited strong melt-freeze hysteresis.